\documentclass[pre,twocolumn,amsmath,amssymb,nofootinbib,floatfix,superscriptaddress]{revtex4}

\usepackage{graphicx,bm}

\makeatletter
  
\begin{document}

\title{ A gauge invariant order parameter for monopole condensation in $QCD$ vacuum. }

\author{ Adriano Di Giacomo} 
\affiliation{ Pisa University and I.N.F.N. Sezione di Pisa}
\email{adriano.digiacomo@df.unipi.it}

\date{\today}

\begin{abstract}
In this paper we improve the existing order parameter for monopole condensation in gauge theory vacuum, making it gauge-invariant from scratch and free of the spurious infrared problems which plagued the old one. Computing the new parameter on the lattice will unambiguously detect weather dual superconductivity is the mechanism for color confinement.

As a byproduct we relate confinement to the existence of a finite correlation length in the gauge-invariant correlator of  chromo-electric field  strengths.

\end{abstract}

\maketitle


\section{Introduction}
Confinement of color is a fundamental problem in particle physics:  hadrons are made of quarks and gluons beyond any reasonable doubt, 
but no free quark or gluon has ever been observed neither in nature nor as a product of a reaction. This phenomenon is known as confinement of color. $QCD$ as the theory of  strong interactions should have the explanation of confinement built in. However at large distances   $QCD$ becomes strongly interacting 
and impossible do deal with the known techniques of field theory, except for numerical simulations on a lattice. The strategy is then to look for "mechanisms", i.e. to explore whether confinement  can  structurally be similar  to some other known physical phenomenon and check this possibility by use of numerical simulations.

A theoretically attractive mechanism for color confinement is dual superconductivity of the vacuum \cite{'tHP} \cite{m}. Confinement of chromo-electric charges is produced by condensation of magnetic charges in the vacuum via dual Meissner effect, in the same way as magnetic charges are confined in ordinary superconductors by condensation of Cooper pairs. 

What makes this mechanism particularly attractive is the fact that it is based on symmetry. The de-confinement  phase transition is a change of symmetry and this provides a "natural" explanation of the fact that the measured upper limits to the existence of free quarks are very small (typically $10^{-15}$ \cite{D}): by this mechanism the number of free quarks in nature is strictly zero by symmetry.

An alternative mechanism suggested in Ref \cite{'tHo} is percolation of center vortices through space-time.

Both mechanisms have been widely studied by use numerical simulations on the lattice. For reviews se e.g. \cite{pol} for monopoles and \cite{eng} for vortices. No conclusive evidence for either mechanism has yet been found.  

The most popular line of investigation of the  monopole mechanism has been to select a gauge such that the corresponding lattice monopoles would dominate the dynamics  of the system (monopole dominance) \cite{suz}. This gauge proves empirically to be  the so called maximal abelian gauge\cite{'tHoo}. The technique was to show that the $U(1)$ system selected by that gauge, and specifically the monopole component of it, would reproduce with good approximation physical observables.

The line of the competing approach based on vortices was very similar: identify a gauge such that the corresponding central vortices would dominate the dynamics  (center dominance). This gauge here  is called the maximal central gauge \cite{ddm}.

Such kind of approaches will never be conclusive: dominance is neither a necessary nor a sufficient condition for monopole condensation or vortex percolation. Indeed it is not obvious at all that one can  describe e.g. a superconducting system by Cooper pair dominance. 

A different  approach is to directly look for   symmetry \cite{Dig}: the vacuum expectation value of an operator\hspace{.2cm} $\mu$ \hspace{.2cm}creating a monopole  can be the order parameter, like the creator of a Cooper pair in an ordinary superconductor. In the superconducting phase the system is a superposition of states with different magnetic charges and $\langle \mu \rangle \neq 0$, in the deconfined phase $\langle \mu \rangle =0$. 

In the $U(1)$ pure gauge theory a construction of the order parameter exists which is rigorous at the level of a theorem \cite{DP} \cite{FM}\cite{PC}. Lattice simulations show that the system has a confined phase at low $\beta $'s [ in the usual notation \hspace{.2cm} $\beta \equiv \frac{2N}{g^2}$ \hspace{.2cm} for gauge group $SU(N)$] and a deconfined phase for $\beta \ge  \beta _c$ , some critical value. The operator which creates a monopole is uniquely identified  as the shift of the vector potential by the classical field of the monopole \cite{DP}, and the shift is generated by the canonically conjugate momentum, the transverse electric field operator. In formulae the operator $\mu (\vec x, t) $ which creates a monopole in the point $\vec x$ at the time $t$ is 
\begin{equation}
\mu(\vec x, t) = \exp ( \int d^3 y \vec E( \vec y, t) \frac{1}{g}\vec A_{\perp} (\vec y - \vec x) ) \label{mono}
\end{equation}
$\frac{1}{g} \vec A_{\perp} (\vec y - \vec x)$  is the vector potential at the  point  $\vec y$ produced by a monopole sitting at $\vec x$, in the transverse gauge. The factor $\frac{1}{g} $ comes from the magnetic charge. An additional factor $\frac{1}{g} $ appears in the lattice formulation, where the electric field is $g$ times the canonical electric field, so that the operator $\mu$ has the form 
\begin{equation}
\mu= \exp( -\beta \Delta S) \label{deltaS}
\end{equation} 

Note that  only the transverse electric field survives the convolution in Eq(\ref{mono}), which is the conjugate momentum to the transverse vector potential in whatever gauge. Lattice simulations \cite{DP} show that $\langle \mu \rangle$ is an order parameter for confinement: 
$\langle \mu \rangle \neq 0$ in the confined phase $\beta \le \beta_c$ and $\langle \mu \rangle = 0$ in the deconfined phase $\beta > \beta _c$ showing that in the $U(1)$ system the mechanism of confinement is dual superconductivity of the vacuum.

The extension of this construction to a generic gauge group  was attempted in subsequent steps by the Pisa lattice group.
The basic difficulty is that the monopole is a $U(1)$ configuration \cite{tH} \cite{Poly}: its existence requires   a Higgs breaking  of the 
gauge symmetry to the $U(1)$ subgroup in
which the Higgs scalar has a definite direction in color space. In $QCD$ there is no Higgs field. Any operator in the adjoint representation can in principle act as an effective Higgs field \cite{'tHoo}, selecting what is called an "abelian projection".
The general attitude of the lattice community was to assume that monopoles belonging to different abelian projections are different objects
and to look for an abelian projection in which monopole dominance were most effective \cite{pol}.

 From the point of view of symmetry instead
the order parameter is the creation operator of a monopole,  and the creation of a monopole should be  a gauge independent process since a monopole  has a non trivial topology. Creating a monopole in any abelian projection amounts to create it in all projections. See on this point Ref \cite{digia} and \cite{digia1}. With this idea in mind the order parameter was  then tentatively constructed for $SU(2)$ gauge group as the creation operator of a monopole in a generic abelian projection [ $U(1)$ subgroup ] specifically along the nominal 3-axis  used in the numerical simulations \cite{ddpp}\cite{dlmp}. $\langle \mu \rangle $ was expected to go to a finite non zero value in the thermodynamic limit $V \to \infty$ below $\beta _c$ and to vanish in the same limit for $\beta > \beta _c$. The approach  looked successful within the lattice sizes available at that time.
Instead an attempt to extend the construction to the gauge group $G2$ \cite{ddd} showed that $\mu$ would tend to  zero in the thermodynamic limit also  in the confined  phase and thus is no order parameter. Also the determinations for $SU(2)$ and $SU(3)$ gauge groups
at larger volumes showed the same problem \cite{ddd}. The origin of the problem was identified and an improved version of the order parameter  was proposed,   infrared subtracted  and numerically tested for  $SU(2)$ pure gauge theory in Ref.\cite{bcdd}.

 In this paper we  elaborate more on the construction of the operator $\mu$ by expanding $\rho \approx \log( \mu)$ , defined in  Eq(\ref{ro}) below,  in a power series of $\Delta S$ i.e. of the magnetic charge of the created monopole [Section II] . The coefficients of the expansion are integrals
 on 3d space or on 4d space of connected correlators of field strengths or of  field strengths squared, which only depend on differences of positions by translation invariance, times functions of the classical field of the monopole which also depend  on the sum of the position vectors.  The correlation functions are exponentially vanishing at large distances in the confined phase and therefore the integrals are finite.
 In the de-confined phase instead there is no length scale, the dependence on the difference of coordinate is dictated by scale dimensions and can give logarithmically divergent  integrals which correspond to zero's of $\mu$, signaling de-confinement. The advantage  of this expansion is a transparent bookkeeping of the terms which can be sensitive to the de-confining transition. [ Section III].

 Moreover the expansion allows to study the divergencies which originate from  the integration on the sum of the positions
: the field  correlators are translation invariant, i.e. they do not depend on the sum of the position vectors.
 Were not for the monopole classical configuration which breaks translation invariance $\rho$  would diverge as $V$ at large volumes $V$. In fact we show that there is only a linear divergence $\propto V^{\frac{1}{3}}$  and only in the first few terms of the expansion, up to $\Delta S^2$. These divergencies which we will call "kinematic" can be isolated  and a regularized $\rho$ can thus be defined to which the  analysis  presented above applies.

 In Section IV we compute  the kinematically divergent part of $\rho$ defined in Section 3  by strong coupling expansion.
 We show that it vanishes to all orders for gauge group $U(1)$, thus confirming the validity of the order parameter, already proved with different arguments in Ref's \cite{DP}\cite{FM}\cite{PC}.
 In \cite{bcdd} it was shown that the strong coupling expansion of $\rho$ is finite to fifth order. 
 
  Divergencies exist instead at that order
 for $SU(2)$ gauge group, they are not even gauge invariant, they are present both in the confined and in the deconfined phase  and are the origin of the problems of the order parameter  observed on the lattice \cite{ddd}. A possible way out is to subtract them by hand \cite{bcdd}. A better way is to improve the definition of the order parameter.
 
 In Section V,  on the basis of the results presented in Section 4, we trace the origin of our problems and modify the order parameter
making  it gauge invariant from scratch, with no "kinematic" divergence, and  conceptually 
 correct. The analyses of sections 2, 3 ,4 are immediately extended to the new operator. 
 
The new operator needs no infrared subtraction. In addition the analysis  allows to connect confinement to the two point gauge invariant correlator of  electric fields:
its contribution to $\rho$ is finite in the confined phase where it decreases exponentially at large distances $x$, but can diverge logarithmically in the deconfined phase where it behaves as $x^{-4}$. We are able to prove in our analysis that this contribution is negative definite, and thus
the divergence corresponds to a zero of the order parameter $\langle \mu \rangle$.

 An overall  discussion of the results is  presented in Section VI.

\section{The order parameter.}
We start analyzing the order parameter as defined in Ref' s \cite{DP}\cite{dlmp}.
 As a consequence of  the proportionality of the monopole field to the inverse of the gauge coupling $\frac{1}{g} $ Eq(\ref{mono}) the order parameter $ \mu \equiv
\langle \mu(\vec x, t) \rangle$ has the form of the ratio of two partition functions \cite{DP} \cite{ddpp}
\begin{equation}
\mu = \frac{Z(S+ \Delta S)}{Z(S)} \label{mu}
\end{equation} 
where  \hspace{.5cm}$Z(S) = \int dA \exp( - \beta S) $ is the $QCD$ partition function.

\vspace{.2cm}
In the ( Wilson) lattice formulation, for pure gauge theory

\begin{equation}
S =  \hspace{.1cm}  \Sigma _{n, \mu, \nu} \Re [\big( P_{\mu \nu} (n) -1 \big)] \label{esse}
\end{equation}

\vspace{.3cm}
$\Re$ denotes  real part,
 $\beta =\frac{2N}{g^2}$ for group $SU(N)$  and $n= (\vec n, t)$ denotes the lattice site. $P_{\mu \nu}(n)$ is the plaquette 
\begin{equation}
P_{\mu \nu}(n) = \frac{1}{N} Tr [ U_{\mu}(n) U_{\nu} ( n + \hat \mu) U^{\dagger}_{\mu}(  n +\hat \nu) U^{\dagger}_{\nu}(n)\big] \label{plaq}
\end{equation}

\vspace{.3cm}
$S+ \Delta S$ has instead the form

\begin{equation}
S+ \Delta S= \hspace{.1cm} \Sigma _{n, \mu, \nu} \Re \big(1 - P'_{\mu \nu} (n) \big) \label{sks}
\end{equation}

with

\begin{eqnarray}
P'_{\mu \nu}(n) = P_{\mu \nu}(n) \hspace{.3cm} everywhere\hspace{.1cm} and \hspace{.1cm}for\hspace{.1cm} all \hspace{.1cm}\mu , \nu \hspace{.1cm}except \nonumber \\
   P'_{i 0} (\vec n, t) = \frac{1}{N} Tr [ U_{i}(\vec n,t) U_{0} (\vec n + \hat i , t) M_{i} (\vec n + \hat i)  \hspace{1.2cm}\nonumber \\ U^{\dagger}_{i}( \vec n , t+1)U^{\dagger}_{0}(\vec n,t)\big] \hspace{2cm} \label{P'}
\end{eqnarray}
Here $t$ is the time at which the monopole is created, which we could fix  at any value by use of  translation invariance.
  
\begin{equation}
M_{i} ( \vec m) = \exp\big ( i g \frac{\sigma _3}{2} \vec {\bar A}_{i} (\vec m -\vec x) \big )  \label{mi}
\end{equation}
$\vec {\bar A}_{i} (\vec y -\vec x)$ is the classical vector potential produced at $\vec y$ by a monopole sitting at $\vec x$ in the transverse gauge $\vec \nabla \vec {\bar A}_{i} (\vec y -\vec x) =0$. It is easily shown by successive changes of variables in the Feynman path integral that replacing $S$ by $S+\Delta S$ is equivalent to add a monopole at the point $\vec x$ in the subgroup $U(1)$ generated by $T_3$ at all times $t > 0$ \cite{dlmp}.

From Eqs(\ref{P'}) and (\ref{mi}) one has
\begin{eqnarray}
P'_{i 0}(\vec n, 0)  =P_{i 0}(\vec n, 0) \cos\big ( \frac{g}{2}\vec {\bar A}_{i} (\vec n + \hat i -\vec x)\big) \nonumber \\+i \sin \big ( \frac{g}{2}\vec {\bar A}_{i} (\vec n + \hat i -\vec x)\big) Q_{i0}(\vec n, 0)  \label{pp}
\end{eqnarray}
where $P_{i 0}(\vec n, 0)$ is defined by Eq(\ref{plaq}) and
\begin{equation}
Q_{i0}(\vec n, 0) \equiv    \frac{1}{N} Tr \big [ U_i (\vec n ,0) U_0 (\vec n + \hat i, 0) \sigma_3 {U^{\dagger}} _i (\vec n, 1) {U^{\dagger}}_0 (\vec n,0)\big]  \label{qi}                                                
\end{equation}

We have assumed gauge group $SU(2)$ for the sake of simplicity. The extension to generic gauge group is straightforward.
  
From Eq.'s (\ref{esse}), (\ref{sks}) and (\ref{P'})  $\Delta S$ is non zero only in the hyperplane  $n_0=t$.  In the following we shall omit the dependence on time, if not specially needed, as well as the dependence on $\vec x$,  the position of the monopole being fixed once and for all. From the definition of $S$ and of $S+\Delta S$ it follows

\begin{equation}
\Delta S = \Sigma _{\vec n} \Sigma _{i}  [ (C_i (\vec n) -1) \Re P_{i0}(\vec n,0) ]- S_i (\vec n) \Im Q_{i0}(\vec n, 0) ]  \label{deltaS}
\end{equation}
To simplify the notation we have  denoted by  $C_i (\vec n)$ and $S_i(\vec n)$ the cosine and the sine appearing in the expression Eq.(\ref{pp}),
leaving the only relevant dependences,  on $\vec n$ and  on $i$. 

Note that the presence of a monopole at $\vec x$  breaks the invariance under spatial translations of $S+ \Delta S$.

 Since $|\vec {\bar A}_{i}(\vec n)| \approx \frac{1}{ n}$ at large distances 
\begin{equation}
(1 -C_i (\vec n) ) \approx \frac{1}{n^2} \hspace{1cm} S_i (\vec n) \approx \frac{1}{n} \label{Ci}
\end{equation}

In the special case $N=1$ the gauge group is abelian, everything commutes, $Q_{i0} =P_{i0}$ , $P_{i0}= \exp{i \theta_{i0}}$. The elementary links have the form
$U_{\mu}(n) = \exp{i\theta _{\mu}(n)}$, $\theta _{\mu \nu}\equiv \theta _{\mu}(n) + \theta_{\nu}(n+ \hat{\mu}) -\theta_{\mu}(n+\hat {\nu}) - \theta_{\nu}(n)$, $P'_{i 0}(\vec n, t)= P_{i0}(\vec n ,t) \exp{ig A_i(\vec n, t)}$ and
\begin{equation}
\Delta S=   \Sigma_{\vec n}  \Sigma_{i} [(C_i(\vec n) -1)  \Re P_{i0}(\vec n, 0)-S_i(\vec n) \Im P_{i0}(\vec n, 0)] \label{deltaSu1}
\end{equation}

To evaluate the order parameter Eq (\ref{mu}) it proves convenient to avoid the direct computation of partition functions, and to compute instead the quantity \cite{DP}. 
\begin{equation}
\rho \equiv \frac{\partial \log(\mu)}{\partial \beta} = \langle S \rangle _S - \langle (S + \Delta S) \rangle _{(S+\Delta S)} \label{ro}
\end{equation} 
The brackets denote average, the subscript on the right brackets denotes the action used to weight the average.
\begin{eqnarray}
\langle S \rangle _S \equiv \frac{\int dA \exp(-\beta S) S}{\int dA \exp(- \beta S)} \hspace{3cm} \nonumber \\ \langle (S + \Delta S) \rangle _{(S+\Delta S)}\equiv\frac{\int dA \exp\big( - \beta (S +\Delta S)\big)\hspace{.2cm}(S + \Delta S)}{\int dA \exp\big( -\beta (S +\Delta S)\big)} \label{spdelta}
\end{eqnarray}
Since $\mu ( \beta=0) =1$
\begin{equation}
\mu(\beta) = \exp \big( \int_0^{\beta}\rho(\beta') d\beta' \big)
\end{equation}
If superconductivity of the vacuum is the correct mechanism for confinement $\rho$ is expected to be finite in the confined phase $\beta < \beta_c $ in the infinite volume limit, so that $\mu \neq 0$. In the deconfined phase $\beta > \beta _c$ $\rho$ must diverge negative in the thermodynamic limit, so that $\mu =0$. This we want to investigate using the definitions Eq(\ref{esse}) and (\ref{sks}) of $S$ and $S+\Delta S$.

\vspace{.3cm}

We shall expand $\langle (S + \Delta S) \rangle _{(S+\Delta S)}$ in powers of $\Delta S$. In our notation\hspace{.2cm}  $ \langle O \rangle \equiv \frac{\int dA  \exp(-\beta S) O}{\int dA  \exp(- \beta S)} $ \hspace{.2cm} for any operator $O$. The expression for $\rho$ Eq(\ref{ro}) is
\begin{equation}
\rho = \langle S \rangle -  \frac{ \langle (S + \Delta S)\big( \sum_{n=0}^{\infty} \frac{(- \beta\Delta S)^n}{n!} \big )\rangle}{ \sum_{n=0}^{\infty} \frac{\langle (- \beta \Delta S)^n \rangle}{n!}} \label{exp}
\end{equation}

\vspace{.3cm}

 $ \Delta S $ is the integral  on 3-d space of an electric plaquette $\Pi _{i0} (\vec n, 0)$ and of  a $Q_{i0}(\vec n , 0)$  multiplied by numerical coefficients,
 and $S$ is in the same way the integral on 4-d space time of  a plaquette times numerical coefficients. Therefore $\rho$ Eq(\ref{ro})
 is expressed in terms of integrals of correlators of plaquettes and ${Q_{0i} }'s$.
 
  We shall first construct an expansion of $\rho$ in powers of $\Delta S$  i.e. in powers of the charge of the external monopole.
  The effect of the series in the denominator of $\rho$ Eq(\ref{exp}) is simply to cancel disconnected parts of the correlators and 
  \begin{equation}
\rho= - \sum_{0}^{\infty} \frac{(- \beta)^{n}}{n!}\langle \langle \Delta s^{n+1}\rangle \rangle -\langle \langle S \sum_{1}^{\infty} \frac{(- \beta )^{{n}}}{n!}  \Delta S^n \rangle \rangle \label{xx}
\end{equation}
  The notation $\langle \langle ..\rangle \rangle$ indicates connected correlators, i.e. correlators with all the disconnected parts subtracted.
  We  prove Eq(\ref{xx}) in  Appendix 1.
  
  As a second basic point of our strategy we note that at the leading infrared order a plaquette is proportional to the square of a component of the field strength, and $Q_{i 0}$ to the  ${i0}$ component of the field strength itself : these terms generate the most infrared part of the correlators, and thus the part of the integrals which can give a divergence at large volumes,
  i.e. a zero of the order parameter $\langle \mu \rangle$ Eq(\ref{mu}). Higher terms will prove to be irrelevant.
  
     The interesting part of $\rho$  will  finally be a sum of gauge-invariant connected correlators of two electric field strengths.  Correlators of gauge invariant fields are known in the literature \cite{dosch} \cite{simonov} \cite{ddss} and have been numerically studied  on the lattice \cite{dp}. We shall come back to this point  below.

The physical idea is that in the confined phase there is a finite correlation length and the correlators are exponentially decreasing at large distances thus making the the integrals infrared convergent, $\rho$ finite and $\langle \mu \rangle\neq 0$ . In the deconfined phase instead there is no length scale in the game, so that the behavior at large distance is dictated by the dimension in length, is power-like and the integral can be negative infrared divergent at large volumes thus making $\langle \mu \rangle =0 $ in the thermodynamic limit $V \to \infty$. This  aspect we  analyze in the next section.

\section{Computing \hspace{.15cm} $\rho$}

 We now analyze in detail Eq(\ref{xx}) both for $SU(2)$ and $U(1)$ gauge groups.  
 
 $U(1)$ will provide a safe test since in that case the order parameter is well defined at a rigorous level \cite{DP} \cite{FM}\cite{PC}.

The generic term $\langle \langle \Delta S^n \rangle \rangle$ is an n-fold 3-d integral of products of factors $\Re P_{i_k0}(\vec n_k)[C_i(\vec n) -1]$ and $ \Im Q_{i0}(\vec n) S_i(\vec n)$. 

Notice that only terms with an even number of factors $Q_{i0}(\vec n)$ survive the average.

This is easily seen  in the case $U(1)$ where \hspace{.1cm}$\Im Q_{i0} \propto \sin( \theta _{i0})$  \hspace{.cm} is odd under the change\hspace{.1cm} $\theta_{\mu} \to -\theta_{\mu}$ \hspace{.1cm}which is a symmetry both of the action $\sum (1-\cos(\theta _{i0})$ and of the measure $\Pi \int_{-\pi}^{+\pi} d\theta _{\mu}(\vec n)$.

For $SU(2)$ gauge group the change of variables

\begin{equation}
U_{\mu}(n) \to \Pi^{\dagger} U_{\mu}(n) \Pi
\end{equation}
 with
\begin{equation}
\Pi \equiv \exp(i \frac{\pi}{2}\sigma _1) \label{pi}
\end{equation}
 inverts the sign of $\sigma _3$ i.e. of $\langle Q_{i0}\rangle $ but leaves  the action and the measure invariant.
 
 For a generic group $\sigma_3$  is replaced by the third component of the $SU(2)$ subgroup in which the monopole lives and the construction is the same.

The generic term of the expansion Eq(\ref{xx}) say $\langle \langle \Delta S^n \rangle \rangle $ will have the form
\begin{equation} 
\langle \langle \Delta S^n \rangle \rangle \propto   \sum_{j_s {\vec n_s}}\Pi _{s=1}^{k}\sum_{i_r {\vec n_r}}\Pi _{r=1}^{n-k}  S_{j_s}(\vec n _s) [C_{i_r}(\vec n_r) -1] F_{i_r  j_s}(\vec n_r, \vec n_s) \label{dimensions}
\end{equation}
The number of factors $\Im Q_{i0}$ is $k$ and must be even.

At the lowest order in the lattice spacing $a$  a plaquette $P_{\mu \nu}$ has the form
\begin{equation}
P_{\mu \nu}  \approx  1 - \frac{a^4}{8} \vec G_{\mu \nu}  \vec G_{\mu \nu}  + .... \label{P}
\end{equation}

The term $1$ does not contribute to connected correlators, the operator in the second term has dimension -4 in length.
 $\Im Q_{io}$ instead has the form  
 \begin{equation}
 \Im Q_{i0} \approx a^2 G^{(3)}_{i0} \label{Q}+ ....
 \end{equation} 
and has dimension -2 in length. The upper index $(3)$ denotes direction in color space.

Higher terms  in the expansion of $P_{0i}(\vec n)$ and $Q_{0i}(\vec n)$ have higher dimension in inverse length.

The function $ F_{i_r  j_s}(\vec n_r, \vec n_s)$ in Eq(\ref{dimensions}) only depends on differences of  $\vec n$ 's by translation invariance and is expected to be cut-off exponentially at large distances in the confined phase, making all the integrals convergent in the infinite volume limit. In the deconfined phase instead there is no intrinsic scale and $ F_{i_r  j_s}(\vec n_r, \vec n_s)$  will depend on inverse powers of the distances with exponent dictated by the scale dimension. Each  factor $\Delta S$ in the correlator contributes  the dimension in length $ l$ of the function as $d^3n[1-C_i(\vec n)] \vec G_{i0}  \vec G_{i0}$, i.e. by  $-3$ [ See Eq(\ref{Ci})]  for the term $P_{0i}$. The insertion of a factor $Q_{i0}$ instead as $d^3nS_i(\vec n)) \vec G_{i0} $ changes the dimension    by $0$.  Therefore terms in Eq(\ref{dimensions}) containing factors
$\Re P_{i0}$ tend to stay finite both in the confined and in the deconfined phase. Terms containing only $Q_{i0}$'s have dimension $0$ and can produce a logarithmic divergence in the deconfined phase.

 Higher terms in the expansions Eq(\ref{P}) and Eq(\ref{Q}) in powers of the lattice spacing $a$ have higher inverse dimension in length and as a consequence they are irrelevant.

As for the terms in Eq(\ref{xx}) $ \langle \langle S \Delta S^n \rangle \rangle$ the result is similar.The insertion of a factor $S$ means, at the leading infrared order, $\int d^4 n
 \frac{a^4}{8} \vec G_{\mu \nu}  \vec G_{\mu \nu}$ which has dimension $0$.

We are tacitly assuming that everything in Eq(\ref{dimensions}) is well defined and finite. 

We immediately realize that this is not the case
by looking at the first few  terms of the expansion  Eq(\ref{xx}) up to $ n=2$. The reason is that the correlators are translation invariant, and only depend on relative distances, but integration on the coordinates also includes  an integral on their sum. Translation invariance is broken by the factors $S_i(\vec n) $  and $(C_i(\vec n) -1)$ containing the field of the monopole. If there are enough of them the integral is convergent [Eq(\ref{Ci})].
By dimensional argument  two factors of type $\Re P_{i0}$ and four factors of type $\Im Q_{0i}$ are sufficient.  In order to understand these  "kinematic" infrared divergencies and get rid of them it will be then sufficient to study the expansion 
Eq(\ref{xx}) up to second order in $\Delta S$, namely
\begin{equation}
\rho \approx -\langle \Delta S \rangle + \langle \langle \beta  S \Delta S \rangle \rangle + \beta \langle \langle \Delta S^2 \rangle \rangle - \frac{1}{2} \beta^2 \langle \langle S \Delta S^2 \rangle \rangle +.... \label{secord}
\end{equation} 
All the other terms are finite.

From Eq(\ref{deltaS}) one easily gets
\begin{equation}
\langle \Delta S \rangle  = \langle \Re P_{i0}\rangle \sum_{i =1 ,3}\sum_{\vec n} (C_i(\vec n) -1)  \propto  \sum_{\vec n} \frac{1}{n^2} \label{O11}
\end{equation}
The term in $\langle \Im Q_{0i}\rangle $ Eq(\ref{deltaS}) vanishes by symmetry. 

$\langle \Re P_{i0}(\vec n)\rangle$ is independent on $\vec n$ due to translation invariance  and on the index $i$ due to invariance under 90 degree rotations around the coordinate axes of the lattice.

 $\langle \Delta S \rangle$ Eq(\ref{O11})  diverges linearly with the  spatial linear size $L$ of the lattice, both in the confined and in the deconfined phase independent of the gauge group.
  
 The term $\beta \langle  \langle S  \Delta S \rangle \rangle$ in Eq(\ref{secord}) contains a term proportional to $P_{0i} $ and a term proportional to $Q_{0i}$. The latter vanishes by symmetry and what is left  is linearly divergent in the infrared

\begin{equation}
\beta \langle  \langle S  \Delta S \rangle \rangle = K  \sum_{\vec n} \sum_{i=1-3} (C_i(\vec n) -1)
\end{equation}
with
\begin{equation}
K = \sum_{n \mu \nu}\langle \langle P_{\mu \nu} (n) P_{0i} (\vec m, t) \rangle \rangle
\end{equation}
$K$ is independent on $\vec m$ and $t$, and is finite in the confined phase but also in the deconfined phase having dimension in length -4.
Note that the integral on the time axis is cut-off by $\frac{1}{T}$ the inverse of the temperature.

 More interesting is the term  $\beta\langle \langle \Delta S^2 \rangle \rangle $ in the expansion Eq(\ref{secord}). By use of Eq(\ref{deltaS}) we get
 \begin{eqnarray}
 \langle \langle \Delta S^2 \rangle \rangle = \sum_{\vec n_1 \vec n_2 i_1 i_2} \big [ \langle \langle \Re P_{0 i_1}(\vec n_1) \Re P_{0 i_2}(\vec n_2) \rangle \rangle (C_{i_1}(\vec n_1) -1) \nonumber \\(C_{i_2}(\vec n_2) -1) +
   \langle \langle\Im  Q_{0 i_1}(\vec n_1) \Im Q_{0 i_2}(\vec n_2) \rangle \rangle S_{i_1}(\vec n_1) S_{i_2}(\vec n_2) \big ] \label{deltaS2}
 \end{eqnarray}
 
 The term $\langle \langle  P_{0i} Q_{0j }\rangle \rangle$ vanishes by symmetry. The first term in Eq(\ref{deltaS2}) is convergent both in the confined and in the deconfined phase: indeed the correlator $ \langle \langle \Re P_{0 i_1}(\vec n_1) \Re P_{0 i_2}(\vec n_2) \rangle \rangle$ has dimension $l^{-8}$ thus making the sum on $\vec n_1-\vec n_2$ convergent in both phases and the factor \hspace{.2cm} $(C_{i_1}(\vec n_1) -1) (C_{i_2}(\vec n_2) -1)$\hspace{.2cm} behaves as 
 $|\vec n_1 +\vec n_2| ^{-4}$ at large distances, thus making the sum on $\vec n \equiv \frac{\vec n_1 +\vec n_2}{2}$ finite. We shall not consider that term any more
 and we shall concentrate on the second term. Indeed adding a finite constant to $\rho$ which is  the logarithm of the order parameter $\langle \mu \rangle$ is equivalent to change the order parameter by a non zero factor and hence it is irrelevant: what matters is that the order parameter be zero or non zero.
 
 We first notice that  $ \langle \langle \Im Q_{0 i_1}(\vec n_1) \Im Q_{0 i_2}(\vec n_2) \rangle \rangle= \langle \Im  Q_{0 i_1}(\vec n_1) \Im Q_{0 i_2}(\vec n_2) \rangle $ since $ \langle Q_{0 i_1}(\vec n)\rangle =0$ by symmetry. Moreover  at the leading infrared order
 \begin{equation}
 \langle \Im Q_{0 i_1}(\vec n_1) \Im Q_{0 i_2}(\vec n_2)\rangle \propto \langle E^{(3)}_{i_1}(\vec n_1) E^{(3)}_{i_2}(\vec n_2) \rangle \label{corrQ}
 \end{equation}
 
$ \vec E^{(3)}_i(\vec n)$ is the $i$ component  of the electric field in color direction 3. For $U(1)$ gauge group the color index can be disregarded.
Invariance under parity implies that the correlator is an even function of $\vec n_1 -\vec n_2$. Moreover the correlator is non zero only if
$i_1 = i_2$. Indeed if one of the electric fields is parallel to $\vec n_1 -\vec n_2$ and the other perpendicular the correlator vanishes due to the invariance under rotations around the direction of $\vec n_1 -\vec n_2$. If they are both transverse to it and perpendicular to each other
a rotation of angle $\pi$ around any of them changes the sign of the other but does not affect the dependence on the distance since  the dependence on it is even.  Therefore we have for the relevant part of $\langle \langle \Delta S^2 \rangle \rangle $ , neglecting terms which are finite both in the confined and in the deconfined phase which are irrelevant to our argument
\begin{equation}
\langle \langle \Delta S^2 \rangle \rangle  \approx  \sum_{i, \vec n_1, \vec n_2} \langle\Im  Q_{0i}(\vec n_1)  \Im Q_{0i}(\vec n_2)\rangle S_i(\vec n_1) S_i(\vec n_2) \label{dssrel}
\end{equation}
This expression diverges linearly when summed on $\vec n = \frac{\vec n_1 + \vec n_2}{2}$ : indeed the correlator does not depend on $\vec n$ and the two factors 
$S_i(\vec n_1) S_i(\vec n_2)$ behave as  $\frac{1}{n}$ each at large distances.  We regularize it and isolate the diverging part by adding and subtracting the contact term  

\begin{equation}
  \Sigma = \sum_{i,\vec n, \vec n_1 - \vec n_2}\langle \Im Q_{0i}(\vec n_1) \Im  Q_{0i}(\vec n_2)\rangle S_i(\vec n) S_i(\vec n)  \propto  \sum_{\vec n} \frac{1}{n^2} \label{DSS}
\end{equation}

 We define a quantity $\langle \langle \Delta S^2 \rangle \rangle _{subtracted}  $ as 
  \begin{equation}
  \langle \langle \Delta S^2 \rangle \rangle _{subtracted}=\langle \langle \Delta S^2 \rangle \rangle  - \Sigma  \label{dssqsub}
 \end{equation}
 
We finally consider the last term of Eq(\ref{secord}) , namely $- \frac{1}{2} \beta ^2 \langle \langle S \Delta S^2 \rangle \rangle$.
As for was for the term $\beta  \langle \langle  \Delta S^2 \rangle \rangle $ of Eq(\ref{deltaS2}) there is a term proportional to $P_{0i_1}(\vec n_1) P_{0i_2}(\vec n_2)$ which is convergent  both kinematically i.e. in the sum over $\vec n$ and by dimension i.e. in the sum over $\vec n_1-\vec n_2$ . We can then  disregard it. As was for Eq(\ref{deltaS2}) the cross term $P_{0i} Q_{0i} $ vanishes by symmetry. We are then left with
\begin{eqnarray}
-  \frac{1}{2} \beta ^2 \langle \langle S \Delta S^2 \rangle \rangle  \approx \frac{1}{2} \beta^2\sum_{n \mu \nu}\sum_{i_1 i_2 \vec n_1 \vec n_2 }S_{i_1}(\vec n_1) S_{i_2}(\vec n_2) \nonumber \\ \langle \langle \Re P_{\mu \nu}( n)  \Im Q_{i_1 0}(\vec n_1) \Im Q_{i_2} (\vec n_2 )\rangle \rangle  
\end{eqnarray}
This expression is "kinematically " divergent both in the confined and in the deconfined phase. As for the other terms of Eq(\ref{deltaS2}) we shall isolate and subtract the divergent part.
By dimensional arguments also this term could be candidate to produce a logarithmic divergence in the deconfined phase: in fact this is not true because the range of the sum on temporal coordinates is cut off  at $\frac{1}{T}$ at non zero temperature in the deconfined phase.

In conclusion the only term of the series defining $\rho$ which can diverge logarithmically in the deconfined phase after removal of the kinematic divergences is $\beta \langle \langle \Delta S^2 \rangle \rangle _{subtracted}$ of Eq(\ref{dssqsub}). 

We notice that  it is negative definite. Indeed starting from the obvious inequality  \hspace{.2cm} $|\sum_{i} S_{i}(\vec n_1) S_{i}(\vec n_2)| \le \frac{  S_{i}^2(\vec n_1) + S_{i}^2 (\vec n_2)}{2}$\hspace{.2cm} and calling $f(\vec n_1 - \vec n_2) \equiv  \langle\Im  Q_{0i}(\vec n_1)  \Im Q_{0i}(\vec n_2)\rangle$ to simplify the notation we have the chain of inequalities
\begin{eqnarray}
 |\langle \langle \Delta S^2 \rangle \rangle |  =|\sum_{i, \vec n_1, \vec n_2}f_i(\vec n_1 - \vec n_2) S_i(\vec n_1) S_i(\vec n_2)| \nonumber \\ \le
 \sum_{i, \vec n_1, \vec n_2}\frac{  S_{i}^2(\vec n_1) + S_{i}^2 (\vec n_2)}{2}f_i(\vec n_1 - \vec n_2) \equiv  \Sigma 
\end{eqnarray}
The last inequality is only true if  $\sum_{\vec n}f_i(\vec n) \ge 0$. Indeed
\begin{equation}
\sum_{\vec n} \langle\Im  Q_{0i}(\frac{\vec n}{2})  \Im Q_{0i}(-\frac{\vec n)}{2})\rangle =\frac{ \sum_{k} |\langle 0 | \Im Q{i0}(\vec 0)|k\rangle| ^2}{\langle k|k \rangle} \label{QQ}
 \end{equation}
 The sum is extended to the states $| k\rangle$ of zero momentum and is certainly positive if the operator $Q_{0i}$ is gauge invariant so that only states of positive metric contribute. We shall come back to this point below.
 
 To summarize we have shown that up to order $\Delta S^2$ $\rho$ is finite, except for a possible logarithmic divergence in the two point correlator of the chromo-electric field. A finite contribution to $\rho$ reflects in a non-zero multiplicative factor in $\mu$ which is irrelevant to symmetry. Higher order terms in the expansion of the plaquettes in terms of lattice spacing  Eq's (\ref{P}) and (\ref{Q}) are then irrelevant, as well as all the terms of order $>2$ in the expansion Eq(\ref{xx}) which contain at least one factor $\Re P_{\mu \nu}$. All the terms which only contain factors  $\Im Q_{0i}$ can diverge logarithmically and be relevant to symmetry.
 We have no idea about the sign of these terms, which could influence the critical index of $\langle \mu\rangle$ at the transition. In the spirit of Stochastic Vacuum \cite{dosch} \cite{simonov} the two point function should dominate. In any case our analysis shows that the order parameter $\langle \mu \rangle$ can be traded  with the inverse of the correlation length of the theory. In Section 6 we discuss  the connection 
 to lattice results on the subject \cite{DMP} \cite{DMP2}.
 We close this section by rewriting the sum of the kinematically divergent parts of $\rho$, $\rho _{div}$ 
\begin{eqnarray}
\rho_{div} = -\langle \Delta S\rangle + \langle \langle \beta S \Delta S\rangle \rangle +\sum_{i,\vec n_1,\vec n_2}[\beta\langle \langle \Im Q_{0i}(\vec n_1) \Im Q_{0i}(\vec n_2)\rangle \rangle \nonumber \\  - \frac{1}{2}\beta^2 \langle \langle S \Im Q_{0i}(\vec n_1) \Im Q_{0i}(\vec n_2)\rangle \rangle ] S_i^2(\vec n)\label{rhodiv} \hspace{0.5cm}
\end{eqnarray}
with $\vec n \equiv \frac{\vec n_1 + \vec n}{2}$

\section{ Computing the kinematic divergences.}

In this section we compute the divergent part of $\rho_{div}$ by use of a strong coupling expansion. It is already known that 
the term $O(\beta^5)$ is convergent for $U(1)$ gauge group, and infrared divergent for $SU(2)$  
 \cite{bcdd}. 

Here we show that for $U(1)$ gauge group  $\rho_{div}$ vanishes to all orders of the strong coupling expansion, as expected \cite{DP} \cite{FM}\cite{PC}.

For  any gauge group and for any operator $O$ the strong coupling expansion of $\langle O \rangle $ is \cite{Creutz}

\vspace{.4cm}
   $\langle O \rangle \equiv \frac{\int \Pi dU_{\mu}(n)  \exp(-\beta S) O}{\int \Pi dU_{\mu}(n)  \exp(-\beta S)}= \sum_{n=0}^{\infty} \frac{(-\beta)^n}{n!}\langle \langle O S^n\rangle \rangle$ \hspace{1cm}.

\vspace{.4cm}
The integral is a group integral, the link $U_{\mu}( n)$ being an element of the group. Here again the double bracket means connected 
graph, the disconnected parts being canceled by the  denominator.

We compute in the strong coupling expansion the four terms in Eq(\ref{rhodiv}):
For the first term  we get $-\langle\langle  \Delta S\rangle\rangle  \equiv  D_1  \sum_{i \vec n} (C_i(\vec n) -1)$, with

\begin{equation}
 D_1 = - \langle \Re P_{i0} \rangle_{SC}= -\sum_{n=0}^{\infty} \frac{(-\beta )^n}{n!} \langle \langle \Re P_{i0} S^n \rangle \rangle  \label{D1}
\end{equation}

Only the term of the action $\sum_{n \mu \nu} P_{\mu \nu}(n) $ Eq(\ref{esse}) contributes to the connected part and not the term $1$.

 Each link must appear an even number of times in the graphs to give a non zero result when integrated over. It follows that only odd values of $n$ are non zero in the sum Eq(\ref{D1}). We rewrite it as
\begin{equation}
D_1 = \sum_{n=0}^{\infty}  \frac{\beta ^{2n+1}}{(2n+1)!} \langle \langle \Re P_{i0} S^{2n+1} \rangle \rangle \label{DD1}
\end{equation}
In the same way we get for the second term $\beta  \langle \langle S \Delta S \rangle \rangle \equiv D_2 \sum_{i \vec n} (C_i(\vec n) -1)$ with $D_2  =\beta \sum_{n=0}^{\infty}   \frac{(-\beta )^n}{n!} \langle \langle \Re P_{i0}S^{n +1}\rangle \rangle $

Here only the even values of $ n$ contribute, but there is an extra $\beta$ with respect to $D_1$  and after some algebra 
\begin{equation}
D_2= \sum_{n=0}^{\infty}  \frac{\beta ^{2n+1}}{(2n)!} \langle \langle \Re P_{i0} S^{2n+1} \rangle \rangle  \label{D2}
\end{equation}

The sum of the two terms gives finally

$-\Delta S + \beta \langle \langle S \Delta S \rangle \rangle = \sum_{i \vec n} (C_i( \vec n) -1) ( D_1+D_2) $
 with
 \begin{equation}
 D_1 + D_2 = \sum_{n=0}^{\infty} \frac{ \beta ^{2n+1}}{(2n+1)!} 2(n+1) \langle \langle \Re P_{i0} S^{2n+1} \rangle \rangle     \label{DD1}
 \end{equation}
 
 We now move to the remaining two terms of Eq(\ref{rhodiv}). We get for the third one [See Eq(\ref{DSS})]  $\Sigma = D_3 \sum_{i \vec n} S_i(\vec n)^2$, and for the fourth term $-\frac{1}{2} \beta ^2 \langle \langle S \Delta S^2 \rangle \rangle  = D_4  \sum_{i \vec n} S_i(\vec n)^2$.
 
 \begin{equation}
 D_3 = -\sum_{n=0}^{\infty} \frac{ (- \beta)^{n+1}}{n !}\sum_{\vec n_2}\langle \langle \Im Q_{i0} (\vec n_1) \Im Q_{i0} (\vec n_2) S^n \rangle \rangle 
 \end{equation}
 
 Here only even values of $n$ contribute ad therefore
 
 \begin{equation}
 D_3 = \sum_{n=0}^{\infty}\frac{\beta^{2n+1}}{(2n)!}\sum_{\vec n_2}\langle \langle \Im Q_{i0} (\vec n_1) \Im Q_{i0} (\vec n_2) S^{2n} \rangle \rangle \label{D3}
 \end{equation}
For the last term we get
\begin{equation}
D_4 =  - \frac{\beta ^2}{2} \sum_{n=0}^{\infty}\frac{ (-\beta)^{n}}{n!}  \sum_{\vec n_2} \langle \langle \Im Q_{i0} (\vec n_1) \Im Q_{i0} (\vec n_2) S^{n+1} \rangle \rangle \label{D4}
\end{equation}
Here only odd values of $n$ contribute, so that 
\begin{equation}
D_4 = \frac{1}{2} \sum_{n=1}^{\infty}\frac{\beta^{2n+1}} {(2n-1)!}            \sum_{\vec n_2}             \langle \langle \Im Q_{i0} (\vec n_1) \Im Q_{i0} (\vec n_2) S^{2n} \rangle \rangle \label{D4}
\end{equation}
Finally for we can write for the sum of the two terms
\begin{eqnarray}
D_3+D_4 = \sum_{n=0}^{\infty} \frac{ \beta ^{2n+1}}{(2n+1)!} (n+1)(2n+1) \nonumber \\\sum_{\vec n_2} \langle \langle \Im Q_{i0} (\vec n_1) \Im Q_{i0} (\vec n_2) S^{2n} \rangle \rangle
\end{eqnarray}
Putting everything together the coefficient of the diverging part , $ D_3 + D_4- \frac{1}{2} (D_1 + D_2)$ is ( the diverging factor of $D_1$, $D_2$
is $C_i(\vec n) -1$, that of $D_3$ and $D_4$ is $S_i^2(\vec n)$ )

\begin{eqnarray}
\sum_{n} \frac{\beta ^{2n+1}}{(2n+1)!} (n+1) \big[ \langle \langle \Re P_{i0} S^{2n+1} \rangle \rangle \nonumber \\  - (2n+1)\sum_{\vec n_2}\langle \langle  \Im Q_{i0}(\vec n_1) \Im Q_{i0}(\vec n_2)S^{2n} \rangle \rangle \big] \label{DIV}
\end{eqnarray}

The analysis above is valid for any gauge group. For the group $U(1)$  $Q_{i0} = P_{i0}$  $\Re P_{i0}$ and $\Im Q_{i0}$ are the real part and the imaginary part of the plaquette and are both gauge invariant. We now show that the divergence cancels to all orders as expected \cite{DP}. For higher groups the problem is not even well defined, since  $\Im Q_{i0}$ is not gauge invariant. In addition there is a divergence at order $\beta ^5$ in the strong coupling expansion,
which is absent in the case of $U(1)$, in agreement with Ref(\cite{bcdd}).

To show that we rewrite the expression in Eq(\ref{DIV}) in a more convenient form:
\begin{equation}
\langle \langle \Re P_{i0} S^{2n+1} \rangle \rangle =(2n+1) \sum_{\vec n_2}\langle \langle \Re P_{i0}(\vec n_1)\Re P_{i0}(\vec n_2) S^{2n} \rangle \rangle \nonumber 
\end{equation}

The choice $\vec n_1$ can be done by use of translation invariance and the factor $(2n+1)$ is purely combinatorial and comes from the exponent of $S^{2n+1}$.

For any value of  $\vec n_1$, $\vec n_2$ and $n$ the contribution to Eq(\ref{DIV}) is proportional to 
  
\begin{eqnarray}
D_n (\vec n_1, \vec n_2) \equiv \langle \langle  \big [\Re P_{i0}(\vec n_1)\Re P_{i0}(\vec n_2) -  \nonumber \\ \Im Q_{i0}(\vec n_1) \Im Q_{i0}(\vec n_2) \big]S^{2n} \rangle \rangle \label{PQ}
\end{eqnarray}
In the case of $U(1)$ gauge group  \hspace{.5cm} $Q_{i0} = P_{i0}$ \hspace{.5cm} and
\begin{eqnarray}
D_n (\vec n_1, \vec n_2)= \frac{1}{4}  \langle \langle  \big [( P_{i0}(\vec n_1)+ P^*_{i0}(\vec n_1)) ( P_{i0}(\vec n_2)  +P^*_{i0}(\vec n_2)) \nonumber \\ +   (P_{i0}(\vec n_1)- P^*_{i0}(\vec n_1))(P_{i0}(\vec n_2)- P^*_{i0}(\vec n_2))\big]S^{2n} \rangle \rangle \hspace{.4cm} \label{PQ}
\end{eqnarray}

The only non zero contributions are those proportional to $P_{i0}(\vec n_1) P^*_{i0} (\vec n_2)$ and $P^*_{i0}(\vec n_1) P_{i0} (\vec n_2)$, and they cancel between the two terms in Eq(\ref{PQ}) so that  $D_n (\vec n_1, \vec n_2)=0$ and there is no divergence. The physical reason is that to have non zero correlation function of two operators they must have total electric flux zero, since the electric flux is odd under charge conjugation. 

Notice that the configurations which survive the group integration $\int^{\pi}_{-\pi} \frac{d\theta_i}{2\pi}$ are those in which each link $U_i(\vec n) =\exp(i \theta _i(\vec n)$  appears the same number of times as its complex conjugate and therefore contributes $1$.

The proof can be made more detailed.We only presented  
 the main point of the argument.We  already know by independent arguments \cite{DP} \cite{FM} \cite{PC}that no kinematic divergence is present
in the order parameter of the $U(1)$ gauge theory. 

To illustrate the argument and to compare to the non abelian case consider the contribution wit $n=2$  i.e. $O(\beta^5)$ depicted in Fig.1
Here $\vec n_2 = \vec n_1 + \hat k $ with k a spatial direction orthogonal to $i$. The two external plaquettes  $P_{i0}(\vec n_1)$ and $P_{i0}(\vec n_2)$ are  depicted by full oriented links, the plaquettes coming from $S^4$ by  dotted oriented  links. The two cubes correspond to the two terms in Eq(\ref{PQ}) and the black dots in the right upper vertices of the external plaquettes represent the $\sigma _3$ insertion in $Q_{i0}$: for  the $U(1)$ case they are trivially 1.

From the graphs it is easy to see that if the links in the external plaquette in $\vec n_1$ rotate clockwise those in the external plaquette in $\vec n_2$ rotate anti-clockwise and vice-versa if we want in each link two lines with opposite direction. This is  an example of the general argument 
given above. For $U(1)$ gauge group there is no $\sigma_3$ insertion, the black dots are equal to $1$  and the two graphs in figure are equal and cancel each other. 

For $SU(N)$ gauge group the graphs can be directly computed by use of the group integration formulae in Appendix 2 \cite{Creutz}.
The result is    $\frac{1}{N^4}$            for the first cube  and zero for the second one. The two contributions do not cancel and the kinematic divergence with them.  This we knew already from Ref.  \cite{bcdd}. The second cube is not even gauge invariant: a generic gauge transformation rotates differently the two $\sigma_3$'s in the plaquette at $\vec n_1$ and in that  at $\vec n_2$
 One way out is to operate a subtraction which eliminates the kinematic divergence but preserves the possibility of connecting a logarithmic divergence of the two-point function of the electric field to the vanishing of the order parameter. This was done in Ref. \cite{bcdd}.
 
 A more satisfactory solution in all respects suggested by  our analysis is  a better definition of the order parameter which is gauge invariant from scratch. This we will discuss in the next section. We show there that in this way the kinematic divergences sum to zero.

\begin{figure}
\includegraphics*[width=0.8\columnwidth]{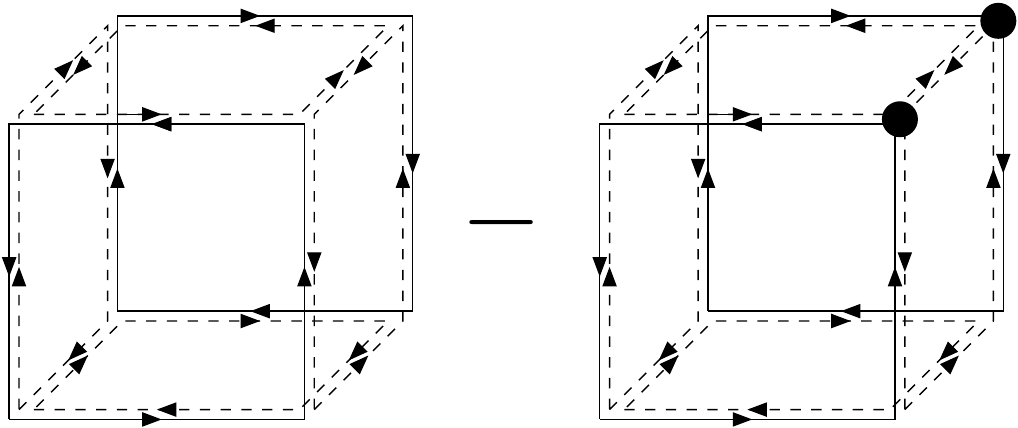}
  \caption{A contribution $O(\beta ^5)$ to $\rho_{div}$. The black circles on the right cube are insertions of $\sigma_3$. }
\end{figure}
\section{A gauge invariant order parameter}
The analysis of Section 4 naturally leads to  a fundamental improvement of the order parameter, which makes it gauge invariant and  free of "kinematic " divergences and sheds light on the meaning of abelian projections.

 To cancel kinematic divergences, i.e. to have $D_n$ as defined by Eq(\ref{PQ}) equal to zero at any order  $n$, $Q_{0i}(\vec n)$  has to be gauge invariant. This looks impossible  since  $Q_{0i}(\vec n)$ transforms as the third component of a vector under local gauge transformations and generically by a different angle in $\vec n_1$ and $\vec n_2$. A change of abelian projection by a local gauge transformation does not help, as well as a non local transformation like a parallel transport to points at finite distance.
 
 A gauge-invariant $\sigma_3$ can however  be defined in any point $(\vec m, t)$ of space-time by parallel transport to infinity,  along any path $C$  by a unitary operator  $V_C(\vec m, t)$ which depends on the point $(\vec m ,t)$ and on the path $C$. We define 
\begin{equation}
 \bar \sigma_3(\vec m, t) = V_C^{\dagger}(\vec m,t) \sigma _3 V_C( \vec m, t) \label{partr}
 \end{equation}
 Any  path $C$ gives a gauge invariant $Q_{0i}(\vec n)$.

 Physically this is procedure is related to the fact that
a monopole breaks some $SU(2)$ symmetry , (generically  a subgroup of the gauge group), to $U(1)$, the little group of  the Higgs field \cite{tH}\cite{Poly}.
This breaking can  not be a breaking of the local gauge symmetry, which is forbidden \cite{elitzur}, but of a global symmetry. The global symmetry lives on the hypersphere at infinity, where the direction $\sigma _3$ is defined, and is gauge invariant.

Indeed any action of the gauge group on a field system has the form \cite{adg}
\begin{equation}
U_G(x)= U( x) U_B 
\end{equation}
where $U(\vec x)$ is the usual gauge transformation at the point $ x$ in the bulk of the system and $U( x) =1$ on the border at infinity where the fields vanish. Instead $U_B=1$ at finite distances, is non trivial at infinity and is a global transformation: it is relevant whenever there are fields which are non zero at infinity, like the Higgs field in the broken phase of a Higgs system.

 We have  shown  that replacing $\sigma_3$ by a parallel transport of it to infinity is a necessary condition to satisfy Eq(\ref{PQ}).
 However  is not generally sufficient as is e.g. in the case for the axial gauge, which is defined by a parallel transport along a line parallel say to the $z$ axis at $\vec x$ and $\vec y$ fixed. It is easily seen that the product of two $Q_i$'s corresponding to two different values of $(\vec x, \vec y)$ is zero to all orders in the strong coupling expansion and thus the second term in Eq(\ref{PQ}) is zero for all values of $\vec n_1$ $\vec n_2$ except 
 for a set of zero measure $(\vec n_1)_x  =(\vec n_2)_x ,(\vec n_1)_y =(\vec n_2)_y$. The second term in Eq(\ref{PQ}) thus vanishes and can not cancel the first term. It is easy to see that the only way to have the second term in Eq(\ref{PQ}) non zero is that the paths $C$ from  different points $\vec n$ to infinity  coincide after some point  $ P$ in their way to infinity: we shall assume that for all the paths $C$ independent of the point. 
 
 We discuss  in detail below the cancellation of $D_n$ i.e. of the kinematic divergence.    Before discussing  the cancellation of the kinematic divergence of the new parameter, we show in detail that it is an order parameter for monopole condensation.
  We have redefined the order parameter by replacing $\sigma _3$ by $\bar \sigma_3(\vec m, t)$ in the expression Eq(\ref{mi}). We get a modified $M_i(\vec m)$ which we call $M_i(\vec m,t)$
  \begin{eqnarray}
  M_i(\vec m,t) = \exp\big ( i g \frac{\bar \sigma_3(\vec m, t)}{2} \vec {\bar A}_{i} (\vec m -\vec x) \big ) \nonumber \\=    V_C^{\dagger}(\vec m,t)M_i(\vec m) V_C( \vec m, t) \label{mipr}
  \end{eqnarray}
We first show that  the new operator creates a monopole as did the old one. Replacing  the action $S$ by $S+ \Delta S$  at any time $t$ is equivalent to create a monopole at all times $>t$, in a similar way as for the old definition. The change of variables in the Feynman path integral \cite{DP}
\begin{equation}
U_i(\vec n, t+1) \to  U_i(\vec n, t+1)M_i (\vec m,t) \nonumber
\end{equation}
leaves the measure invariant and sends the quantity $P'_{i0}(\vec n, t)$ of Eq(\ref{P'}) to $P_{i0}(\vec n , t)$, the same as with the old definition.
The spatial links appearing in the magnetic plaquettes at time $t+1$ get modified as
\begin{equation}
U_i (\vec n, t+1) \to M_i(\vec m,t) U_i (\vec n, t+1)  \nonumber
\end{equation}
which means that a monopole has been added in the color direction $\bar \sigma_3(\vec m, t)$ or in the gauge invariant direction $\sigma _3$ on the sphere at $\infty$. The old definition would add it in the color direction of the $\sigma _3$ axis.
Finally the link affected by the change of variables appears in the plaquette $P'_{i0}(\vec n, t+1)\equiv P_{i0}(\vec n, t+1)$ which is changed to
\begin{eqnarray}
P'_{i0} (\vec n , t+1) = \frac{1}{N} Tr \big[ U_i (\vec n, t+1) U_0 (\vec n+\hat i ,t+1) \nonumber \\ {U^{\dagger}}_0 (\vec n+\hat i ,t+1)M_i (\vec n + \hat i,t) U_0 (\vec n+\hat i,t+1)\nonumber \\{U^{\dagger}}_0(\vec n, t+2)U^{\dagger}_0 (\vec n, t+1)\big]
\end{eqnarray}

We can define 

\begin{equation}
M_i (\vec n + \hat i,t+1) = {U^{\dagger}}_0 (\vec n+\hat i ,t)M_i (\vec n + \hat i,t)U_0 (\vec n+\hat i,t)  \label{redef}
 \end{equation}
 or 
 \begin{equation}
 V_C( \vec n, t+1)=V_C( \vec n, t) U_0(\vec n,t) \label{newC}
 \end{equation}
 The net effect of the change of variables has been to expose the monopole at time $t+1$ and to reproduce at time $t+1$ the same situation that existed originally at time   $t$ with the new path of the form Eq(\ref{newC}). 
 
  Iterating the change of variables proves our statement. 
  
  Comparing to the old definition the expression for $\Delta S$ Eq(\ref{deltaS}) stays unchanged with the same $\Re P_{i0}(\vec n)$ in the first term but a modified $Q_{i0} (\vec n)$ with respect to the one defined in Eq(\ref{qi}): in the new definition $\sigma_3$ is replaced by $\bar \sigma_3(\vec m, t)$
  \begin{equation}
 Q_{i0}(\vec n,t) = \frac{1}{N} Tr \big[ U_i(\vec n, t) U_0(\vec n + \hat i,t)\bar \sigma _3 U_i^{\dagger} (\vec n, t+1)U_0^{\dagger} (\vec n,t) \big]  \label{qip}
  \end{equation}
  The new $Q_{i0}$ is a gauge invariant electric field strength.
As for  the old one the vacuum correlator of an odd number of  $Q_{i0}$'s is zero. Indeed if we replace each link  $U_{\mu}(n)$ in the Feynman integral  by $\Pi U_{\mu}(n) \Pi^{\dagger}$ with $\Pi =\exp(i \sigma_1\frac{\pi}{2})$, the action and the measure stay invariant but $\bar \sigma_3(\vec m, t)$ changes sign. 

The analysis of Section 3 based on dimension in length of the correlation functions stays unchanged. But now the two point function
 $\langle \langle \Im Q_{i0}(\vec n_1,0) \Im Q_{i0}(\vec n_2,0)\rangle \rangle$ is the gauge invariant connected correlator of two electric fields. Such quantities have been studied in  the literature \cite{dosch}\cite{simonov}
\cite{ddss},in particular on the lattice \cite{dp}. Its subtracted version Eq(\ref{dssqsub}) is the candidate part to signal de-confinement 
by diverging logarithmically  at the phase transition, by pure dimensional arguments. With the new definition that quantity is gauge invariant and well defined and the proof  that it is negative definite Eq(\ref{QQ}) is valid, being $Q_{i0}$ gauge invariant.

Finally we  argue  that the new parameter has no kinematic divergence. The terms $D_3$ and $D_4$ of Section 4 are now well defined as in $U(1)$ gauge theory and not gauge dependent. We show that they cancel with $D_1$ and $D_2$ like in the $U(1)$ theory order by order in the strong coupling expansion. 

 To be definite consider in the case of the two point function $\langle \langle Q_i(\vec n_1,0)Q_i(\vec n_2, 0)\rangle \rangle$  for the  path $C$ a straight line from $(\vec n_1, 0)$ $(\vec n_2,0)$ to $(\frac{\vec n_1+\vec n_2}{2}, 0)$ and then a common path along some axis to infinity [Fig(2)]. It is easily seen[Appendix 2] that
 the result is a gauge invariant connected two point function of electric field strengths as defined in Ref  \cite{dp}: See Fig 2.  
 \begin{figure}
\includegraphics*[width=0.8\columnwidth]{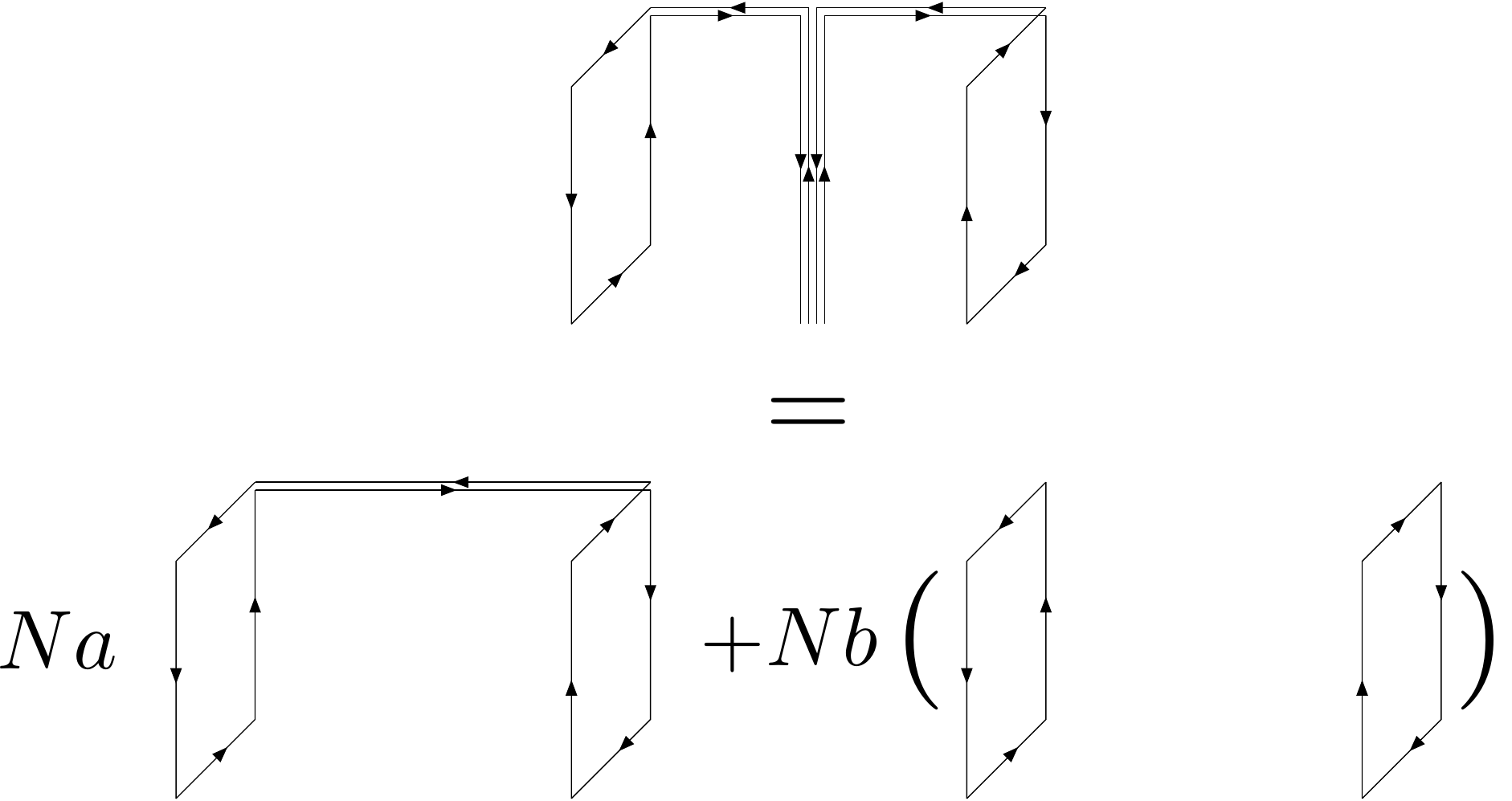}
  \caption{The product of two gauge-invariant fields.\\ $ a= \frac{1}{N^2-1},\hspace{.5cm} b= -a \frac{1}{N}$. The proof in Appendix B}
\end{figure}
The meaning of the equality in Fig.2 is that the graphs represented can be parts of a generic configuration to be integrated over the links in the strong coupling expansion of any correlation function: only the integral on the four overlapping links along the time axis has been performed. The result is independent  of the  line transporting to $ \infty$ : the product of any number of tensors of the form Eq(\ref{C4})is a tensor of the same form, i.e. that tensor is a projector.  If we choose a different path for the parallel transport the result is a connected two point correlator with the connecting line of different form: the only condition is that the two paths originating from $Q_{i0}(\vec n_1)$ and $Q_{i0}(\vec n_2) $ overlap at some point on their way to $\infty$. All this game can be repeated for an n-point function of gauge invariant chromo-electric fields but we will not do that since higher correlators are not relevant to the diverging part of $\rho$ and we shall neglect their contribution  to $\rho$ itself in the spirit of the stochastic vacuum model \cite{dosch} \cite{simonov}. To compare to the old approach, whenever we have two $Q_{i0}$'s in a term of the strong coupling expansion, which were represented as plaquettes with a black dot  we have to replace them by the expression in Fig.2, and then perform the integrations on the links.
 In Appendix 2   we do that explicitly for the right cube in Fig.1 with the result that it exactly cancels with the left cube. The general procedure is to integrate first on all the links different from those appearing in Fig.2. If  after that the parallel transport between the two fields acquires no overlapping links the result is zero because of Eq(\ref{C2})  and of the fact  that $b=-\frac{1}{N}a$.
If instead two extra links  are left overlapping with the parallel transport in Fig.2 it is easily shown by use of the result in Fig.2 and of Eq(\ref{C4})  that the term cancels exactly the corresponding term coming from two $P_{0i}$'s.
In principle one could expect that also terms exist for which more than one pair of extra  links appear,
and one should extend the proof to them. We shall not do that here and assume that cancellation as a  natural conjecture for the time being.

\section{Discussion}

The order parameter for monopole condensation is  $\langle \mu \rangle$ the $vev$ of the creation operator of a monopole. $\mu$ is  the shift by the classical field of a monopole of the transverse vector potential operated by use of the conjugate momentum, which is the transverse electric field. This is basic quantum mechanics. If dual superconductivity is the mechanism for confinement we expect   $\langle \mu \rangle\neq 0$ in the confined phase and  $\langle \mu \rangle=0 $ in the deconfined one. This is exactly what happens in the $U(1)$ gauge theory on the lattice \cite{DP} \cite{FM} \cite{PC}.

In the non abelian case, say $SU(N)$ one would naively expect that the order parameter is the $vev$ of the operator which creates a monopole in some $U(1)$ subgroup of the gauge group. The argument is that creating a monopole is a gauge invariant operation, since the monopole is a configuration with non trivial topology \cite{digia1}: as a consequence the specific choice of a $U(1)$ subgroup should be irrelevant \cite{dlmp}. The result of this procedure, however, proves to be a nonsense: the resulting order parameter vanishes in the thermodynamic limit $V \to \infty$  both in the confined and in the deconfined phase and therefore is no order parameter \cite{bcdd}.

In this paper we have analyzed in detail the structure of the order parameter for generic gauge group, by expanding the quantity $\rho$ Eq(\ref{ro}) in powers of $\Delta S$ [Eq(\ref{sks}) and (\ref{deltaS})]. We found that the first few terms of the expansion are divergent at large volumes independent of the  gauge group and of the dynamics of the gauge theory. We have  isolated these divergences which we call kinematic divergences.  We show that for gauge group $U(1)$ they cancel among themselves, so that the order parameter is well defined. For non abelian theories instead we have tracked the origin of our problems in the fact that the kinematic divergences do not cancel, $\rho$ diverges in the thermodynamic limit both in the confined and in the deconfined phase thus spoiling the possibility of $\langle \mu \rangle$ of being an order parameter. In addition the divergent part is not even gauge invariant.
This indicates that there is something deeply wrong in the procedure. Indeed a monopole breaks some $SU(2)$ symmetry to $U(1)$ and that $SU2)$ can not be a gauge symmetry but only a global symmetry, e.g. a group $SU(2)$ at infinity. The field strengths are to be replaced by gauge invariant field strengths \cite{dosch} \cite{simonov} \cite{ddss}.  As a byproduct this makes the kinematic divergence zero and the order parameter well defined and gauge invariant.

What is left of the order parameter after the cancellation of the kinematic divergences is finite in the confined phase if the correlation functions of the field strengths are exponentially cut-off at large distances as is in presence of a mass gap. The only way to have $\langle \mu \rangle \to 0$ at the deconfining transition is that $\rho \approx _{V \to \infty} K \ln(V)$ with $K < 0$ some constant. The value of $K$ is related to the critical index $\delta$ by which
$\langle \mu \rangle \to 0 $ at  the deconfining temperature $T_c$. We identify all the terms in our expansion which can in principle have such a behavior: they are all the correlators which do not contain $P_{i0}$'s  (plaquettes or field strengths squared) but only $Q_{i0}$'s ( field strengths). In particular we show that the two point function of gauge invariant electric fields is negative definite and can produce a zero of $\langle \mu \rangle$. In the spirit of the stochastic vacuum model \cite{dosch} \cite{simonov} this term should dominate.
Attempts exist in the literature to determine numerically its  behavior as a function of the temperature \cite{DMP} \cite{DMP2}. A precise determination of $K$ and a comparison to the measured value of $\delta$ could in principle say something on the contribution of higher correlators i.e. on the validity of the stochastic vacuum model.
In any case our analysis relates confinement to the existence of a finite length at least in the correlation of chromo-electric fields. This subject
is also studied in different approaches  [ see e.g. Ref. \cite{SC}].

Finally a comment about the uniqueness of the order parameter $\langle \mu\rangle$. The direction of the common path to infinity is irrelevant by symmetry reasons, as well as the position of the point $P$ on it. Different choices for the path before the point $P$ lead to a different line of parallel transport between the two points. As long as for all of them there is a finite correlation length in the confined phase and not in the deconfined phase, they all produce the same order parameter. Indeed adding a finite number to $\rho$ which is roughly the logarithm of $\langle \mu \rangle$, results in a non zero multiplicative constant for the order parameter which we have considered irrelevant in the whole analysis of this paper, since it does not affect the fact that it is zero or non zero.

The possibility should also be studied of computing the new gauge-invariant $\rho$  on a lattice. A choice for the implementation could be to have all the parallel transports go to a point, say the origin of spatial coordinates and then  to $\infty$ as in Fig.2. 
This research would  clarify in an unambiguous way whether dual superconductivity of the vacuum is the correct mechanism for confinement.

\section{Appendix 1}
We want to prove Eq(\ref{xx}) .

To do that we first compute the series expansion of the quantity $\frac{1}{D}$ with $D$ the denominator in Eq(\ref{exp}).
\begin{equation}
D = \sum_{n=0}^{\infty}\frac{\langle (- \Delta S)^n \rangle}{n!} \label{D}
\end{equation}
The result is
$\frac{1}{D} = 1 + \sum_{n=1}^{\infty} d_n$ with
\begin{eqnarray}
d_n = \frac{(-)^{n+1}}{n!} [ \langle \Delta S ^n \rangle - \sum_{k_1=1}^{n-1} \frac{n!}{k_1 ! (n-k_1)!} \langle \Delta S \rangle ^{k_1} \langle \Delta S \rangle ^{n- k_1}\hspace{6cm}\nonumber  \label{dn} \\+ \sum_{k_1\ge1,k_2\ge1} \frac{n!}{k_1! k_2 ! (n-k_1-k_2)!} \langle\Delta S^{k_1}\rangle \langle \Delta S ^{k_2}\rangle \langle \Delta S ^{n-k_1-k_2} \rangle \nonumber \hspace{5.8cm} \\ -.....+(-)^{n+1} \frac{n!}{1!^n} \langle \Delta S\rangle ^n ] \label{dn}\hspace{9cm}
\end{eqnarray}

The sums over the $k_i$'s in all terms run on positive integers with the condition $n - \sum k_i \ge 1$. The last term is the on in which all of the $k_i= 1$, $( i=1...n-1)$ and   $n - \sum k_i = 1$. 

According to the definition of connected correlator  
\begin{equation}
d_n = \frac{(-)^{n+1}}{n!} \langle \langle \Delta S^n \rangle \rangle \label{dnn}
\end{equation}
Indeed the expression in Eq(\ref{dn})   subtracts all the disconnected parts from the correlator  $\langle \Delta S^n \rangle$.

\vspace{.5cm}
We then compute the series expansion of the quantity  $T\equiv \frac{1}{D} \exp( -\Delta S)$ which appears  in  Eq(\ref{xx}) . 

We get $T= \sum_{n=0}^{\infty} T^{(n)}$.  $T^{(0)}=1$ and for $n \ge 1$ 
\begin{equation}
T^{(n)} = \sum_{k=0}^n \frac{ (- \Delta S)^{n-k}}{(n-k)!} d_k
\end{equation}
or, isolating the term with $k=n$,
\begin{equation}
T^{(n)} = \frac{(-)^n}{n!} \bar  {\Delta S^n} + d_n \label{Tn}
\end{equation}
where
\begin{eqnarray}
  \bar {\Delta S^n} \equiv \Delta S ^n  - \sum_{k_1=1}^{n-1} \frac{n!}{k_1 ! (n-k_1)!}  \Delta S ^{k_1} \langle \Delta S \rangle ^{n- k_1}\hspace{6cm}\nonumber \\+ \sum_{k_1\ge1,k_2\ge1} \frac{n!}{k_1! k_2 ! (n-k_1-k_2)!} \Delta S^{k_1} \langle \Delta S ^{k_2}\rangle \langle \Delta S ^{n-k_1-k_2} \rangle \nonumber \hspace{6cm}\\ -.....+(-)^{n+1} \frac{n!}{1!^n} \Delta S \langle \Delta S\rangle ^{n  -1}\hspace {8cm}
\end{eqnarray}

It is immediately seen that

$\langle \bar {\Delta S^n} \rangle =\langle \langle \Delta S^n \rangle \rangle$

Moreover  by use of Eq(\ref{dnn}) and Eq(\ref{Tn})$\langle T^{(n)}\rangle =0$  so that  $\langle T \rangle =T^{0} =1 $ as it should be.

 $ \bar {\Delta S^n} $ is the connected part of  $\Delta S^n$.
 If $O$ is any local operator the quantity
 
 \begin{equation}
 \langle O T^{(n)} \rangle = \frac{(-)^n}{n!}\big[ \langle O \bar \Delta S^n \rangle - \langle O \rangle \langle \langle \Delta S^n \rangle \rangle \big]=\langle \langle O \Delta S^n \rangle \rangle
 \end{equation}
 is fully connected being the connected part  of the correlator of $O$ with a connected correlator.
 Taking $O =S+\Delta S$  proves Eq(\ref{xx}). 
 
 The result for the term proportional to $\Delta S$, $T_2$ is known in the literature \cite{vk}.
 
 Indeed 
 \begin{equation}
 T_2 = \frac{ \sum_{n=0}^{\infty}{\frac{ \langle (-\Delta S )^{n+1}\rangle}{n!}}}{\sum_{n=0}^{\infty}\frac{ \langle  (-\Delta S )^n\rangle}{n!}}
 \nonumber 
  = \partial_{\lambda} \ln( \langle \exp (-\lambda \Delta S)\rangle)_{\lambda=1} \nonumber
  \end{equation}
 or \cite{vk} 
\begin{equation}
T_2 = \sum_{m=0}^{\infty}  \frac{(-)^m}{m!} \langle \langle \Delta S ^{m +1}\rangle \rangle 
\end{equation}

The logarithm  of a generating functional is the generator of the connected correlators, a well known fact.

\section{Appendix 2}

We make  use in our strong coupling computations of two basic formulae which we take from Ref.\cite{Creutz}. The first one is
\begin{equation}
\int dU U_{\alpha _1 \beta_1} U^{\dagger}_{\beta_2 \alpha_2}= \frac{1}{N}\delta_{\beta_1 \beta_2} \delta_{\alpha_1 \alpha_2} \label{C2}
\end{equation}
 The group is $SU(N)$, $U$ is an $N\times N$ matrix in the fundamental representation, and the integration 
ranges on the group.

The second basic formula is 
\begin{eqnarray}
\int dU U_{\alpha_1 \beta_1}U^{\dagger}_{\beta_2 \alpha_2}U_{\alpha _3 \beta _3} U^{\dagger}_{\beta_4 \alpha_4} = \hspace{2cm}\nonumber \\ a [ \delta _{\alpha_1 \alpha_2} \delta_{ \alpha_3 \alpha_4} \delta_{\beta_1 \beta _2} \delta_{\beta_3 \beta_4} +\delta_{\alpha_1 \alpha_4} \delta_{\alpha_2 \alpha_3}  \delta _{\beta_1 \beta_4} \delta_{\beta_2 \beta_3}] +\nonumber \\ b [\delta _{\alpha_1\alpha_2} \delta_{ \alpha_3 \alpha_4} \delta _{\beta_1 \beta_4} \delta_{\beta_2 \beta_3} + \delta_{\alpha_1 \alpha_4} \delta_{\alpha_2 \alpha_3}\delta_{\beta_1 \beta _2} \delta_{\beta_3 \beta_4} ] \label{C4}\hspace{.5cm}
\end{eqnarray}
with 
\begin{equation}
\hspace{2cm}a = \frac{1}{N^2-1}\hspace{1.0cm}   b =- \frac{1}{N} a \label{ab}
\end{equation}

A first consequence of Eq(\ref{C2})  is that the average value of any closed path covered by two lines circulating in opposite direction   is equal to 1. This allows to immediately compute the left cube in Fig.1: the integral on the horizontal pairs of links connecting the front and the rear plaquette gives by Eq(\ref{C2}) $\frac{1}{N^4}$ times the product of the two plaquettes covered each by two lines circulating in opposite directions, which is $=1$. In conclusion the $vev$ of the left cube is  $\frac{1}{N^4}$ . 

As for the cube on the right in Fig.1
we can repeat the procedure, but now of the two overlapping lines both in the front and in the rear plaquette one contains a $\sigma_3$ inserted and when the average is taken by use of Eq(\ref{C2}) the result is proportional to $ (Tr \sigma_3)$ and thus is zero.
The difference of the two cubes is non zero and with it the kinematic divergence.

We now prove the equality in Fig.2 . We integrate on the four overlapping links in the central vertical line by use of Eq(\ref{C4}). Of the four terms two
are proportional to $[Tr \sigma_3]^2$ and vanish, the other two are proportional to $Tr(\sigma_3)^2 = N$ which is the factor in front of the result. The coefficients $a$ and $b$ are computed in Ref \cite{Creutz}. We could have extended arbitrarily the length of the central line      down to $\infty$, with the same result. It is indeed easy to show that the product of two tensors of the form Eq(\ref{C4}) has the same form with the same coefficients $a$ and $b$. We could also have chosen a different form of the paths merging in a single line after some point on the way to infinity, as well as a different direction to infinity.The result would only be 
a different path for parallel transport connecting the two plaquettes  in Fig.2. Notice that the equality in Fig.2  means that the fields 
there can be part of a generic  configurations to integrate over. The only integral which has already been performed is that on the  link in which the lines merge. To have the equality in Fig.2 an exact equality it is necessary to extend the line to infinity. At any finite order of the strong coupling expansion the connected contributions to any correlation function extend to a finite distance, which, however can tend to infinity with increasing order.

Finally we discuss the cancellation of the kinematic divergence. We start computing the right cube in Fig.1. With the gauge-invariant order parameter the two external  $Q_{i0}$'s  are replaced by the two terms in the right hand side in Fig.2. Integrating on the double links between the front and the rear plaquette by use
of Eq(\ref{C2}) and then on the fourfold link by use of Eq(\ref{C4}) gives after some algebra $\frac{1}{N^4}$ multiplied by the quantity
\begin{equation} 
a N^2[ a(N^2+1) + 2 bN ] + bN=  a(N^2-1) =1
\end{equation}
The two cubes are equal and the divergence cancels exactly.

This is not only true for the cubes in Fig.1 but for all the pairs of configurations in which two plaquettes of the first are replaced by two
$Q_{i0}$'s  i.e. by the path in Fig.2 in the second one.  After integration on all the links except those of that path itself  whenever one is  left 
with one  additional pair of opposite links like in the case of the cubes of Fig.1  the cancellation works, independent of the distance between the external plaquettes and on the shape of the parallel transport between them: this is a consequence of the fact that the tensor Eq(\ref{C4}) is a projector.  One should prove that the cancellation also works when after integration  one or more additional pairs of links appear superimposed to the original parallel transport between the two $Q_{i0}$'s. We will  not do that here.

If instead there is no extra pair of links superimposed to the parallel transport the integration on the original links gives $a \frac{1}{N}$ times the disconnected term, and this cancels with the term proportional to $b$. The configuration is disconnected.

\end{document}